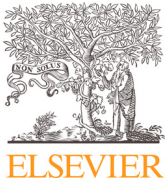
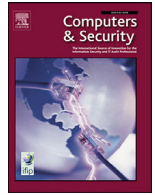

# Perspectives of non-expert users on cyber security and privacy: An analysis of online discussions on twitter

Nandita Pattnaik[1,]*, Shujun Li[2,]*, Jason R.C. Nurse[3]

*Institute of Cyber Security for Society (iCSS) & School of Computing, University of Kent, Canterbury, CT2 7NP, UK*

**A B S T R A C T**

Many researchers have studied non-expert users' perspectives of cyber security and privacy aspects of computing devices at home, but their studies are mostly small-scale empirical studies based on online surveys and interviews and limited to one or a few specific types of devices, such as smart speakers. This paper reports our work on an online social media analysis of a large-scale Twitter dataset, covering cyber security and privacy aspects of many different types of computing devices discussed by non-expert users in the real world. We developed two new machine learning based classifiers to automatically create the Twitter dataset with 435,207 tweets posted by 337,604 non-expert users in January and February of 2019, 2020 and 2021. We analyzed the dataset using both quantitative (topic modeling and sentiment analysis) and qualitative analysis methods, leading to various previously unknown findings. For instance, we observed a sharp (more than doubled) increase of non-expert users' tweets on cyber security and privacy during the pandemic in 2021, compare to in the pre-COVID years (2019 and 2020). Our analysis revealed a diverse range of topics discussed by non-expert users, including VPNs, Wi-Fi, smartphones, laptops, smart home devices, financial security, help-seeking, and roles of different stakeholders. Overall negative sentiment was observed across almost all topics in all the three years. Our results indicate the multi-faceted nature of non-expert users' perspectives on cyber security and privacy and call for more holistic, comprehensive and nuanced research on their perspectives.

## 1. Introduction

People's use of the internet has been increasing at a very fast pace in the past decades. A recent report from Statista (Johnson, 2022b) informed that roughly 62% of the world population and 94% in Western Europe were using the internet as of January 2022. In terms of computing devices they use to access the internet, another 2021 report from Statista (Johnson, 2022a) found that almost 91% of global users used smartphones, followed by 71% using laptops and/or desktop PCs, 30% from smart TVs, and 14% from other smart home devices. The same report also showed that most (65%) of global users accessed the internet from their own computing devices, while a significant portion (30%) used work devices. With the wide use of the internet and different types of computing devices, cyber security and privacy issues have also been increasing, and many internet users have developed some awareness of such issues, especially on phishing and malware (Johnson, 2021a).

With the increasing use of the internet and computing devices by global users, and the consequent cyber security and privacy issues, it is important to understand the perspectives of users regarding such topics. *User perspective* in this context includes the levels of awareness, perception, concerns, decisions, behaviors, and the day-to-day practice of the users in relation to cyber security and privacy. Many researchers have conducted studies to better understand users' perspectives on cyber security and privacy. Related studies have mainly focused on two different types of users – 'experts' with a good level of cyber security knowledge and 'non-expert users' who normally lack cyber security awareness (Rahman et al., 2021). However, most past studies Kang et al. (2015); Wu and Zappala (2018) primarily relied on data self-reported by a relatively small number of recruited human participants, and often qualitative methods were used due to the small scale of the empirical studies (Rahman et al., 2021). Results obtained from such smaller-scale self-reported data are very useful but normally cover only a narrow range of topics, and key findings can be difficult to generalize to a larger population of users. Therefore, working with larger-scale data collected from the wild, e.g., real-world data collected from online social network (OSN) platforms, can help provide more evidence and insights supplementary evidence to past studies (Andreotta et al., 2019). However, filtering out specific cyber security and privacy related data, especially for non-expert users, from the massive number of online posts is a challenging task not only because of the sheer volume, but also because it is not trivial (1) to filter out cyber security and privacy related posts as non-expert users will not always use cyber security or privacy related keywords in their posts and (2) to decide whether the author of a post is an expert or a non-expert user or an organization.

* Corresponding authors.
   *E-mail addresses:* np407@kent.ac.uk (N. Pattnaik), S.J.Li@kent.ac.uk (S. Li), J.R.C.Nurse@kent.ac.uk (J.R.C. Nurse).
   [1] ORCID: 0000-0003-1272-077X
   [2] ORCID: 0000-0001-5628-7328
   [3] ORCID: 0000-0003-4118-1680

Furthermore, there has been a general trend of the focus on the computing devices covered, from more traditional ones such as PCs and smartphones (Camp, 2009; Kang et al., 2015; Wash, 2010) to smart and IoT (Internet of Things) devices more recently (Barbosa et al., 2020; Cannizzaro et al., 2020; Chalhoub and Flechais, 2020; Sturgess et al., 2018; Zheng et al., 2018; Zimmermann et al., 2019). Most related research in the literature arguably focused more on one type of computing device (e.g., smartphones or smart speakers), therefore does not cover the wider range of computing devices modern users are using in different contexts of cyber security and privacy.

We, therefore, decided to do a large-scale analysis of cyber security and privacy related user discussions on social media, especially those posted by non-expert users that are related to different types of computing devices (both smart and traditional ones) used in a home networking environment. This paper reports our analysis of nearly half a million tweets about non-expert users' discussions on Twitter, one of the largest OSN platforms, to provide a more holistic view of their perspectives on cyber security and privacy. We chose to focus on non-expert users in our study, since the rich knowledge and different mental models of cyber security experts make them a less representative user group to study normal users' perspectives (Busse et al., 2019; Camp et al., 2008). The dataset covers three 2-month periods in three years (January and February in 2019, 2020 and 2021), allowing us to look at yearly trends as well as changes before the first global COVID-19 lockdowns and during the COVID-19 pandemic. More specifically, our research questions (RQs) are defined as follows:

RQ1 How can we identify cyber security and privacy related tweets posted by non-expert users?
RQ2 What do non-expert users normally discuss online, regarding their personal perspectives on cyber security and privacy?
RQ3 How did such discussions on Twitter evolve in the past three years, including before and during the COVID-19 pandemic?
RQ4 What was the general sentiment of non-expert users when talking about cyber security and privacy on Twitter?
RQ5 To what extent did the general sentiment of non-expert users change in the past three years, including before and during the COVID-19 pandemic?

Based on the above RQs, our main contributions can be summarized below:

1. We designed a machine learning based classifier to detect specific tweets related to cyber security and privacy with a good performance. To the best of our knowledge, past studies have focused on topics such as classification of cyber security related accounts (Mahaini et al., 2019) or cyber threats (Dionísio et al., 2019).
2. We also developed a second machine learning based classifier to detect non-expert user accounts with a good performance. We were unaware of any work on such classifiers.
3. By applying topical modeling to our dataset, we identified a wide range of topics non-expert users discussed on cyber security and privacy, such as on VPNs, Wi-Fi, smartphones, laptops, smart home devices, financial security, and different stakeholders.
4. A trend analysis of non-expert users' discussions on cyber security and privacy across the three years showed some noticeable changes of topics, e.g., a substantial increase of VPN use during the COVID-19 pandemic compared with before it.
5. By applying sentiment analysis to the datasets and tweets belonging to different topics, this study noticed a general negative sentiment of non-expert users when discussing cyber security and privacy on Twitter.
6. A trend analysis of non-expert users' sentiment across the three years revealed some previously unknown patterns, e.g., the sentiment seemed to have become more polarized over years, with the percentage of more neutral posts decreasing year on year.

The rest of the paper is organized as follows. After briefly introducing some related work in Section 2, the methods we used for this study are explained in detail in Section 3. Section 4 reports results and findings of the research, followed by Section 5 in which further discussions, limitations of the research and our future work are presented. The last section concludes the paper.

## 2. Related work

With the growing level of social media activities, online users have been generating massive textual content (Andreotta et al., 2019), and researchers have been increasingly utilizing such user-generated content (UGC) to study different online phenomena (Yin and Kaynak, 2015). However, to the best of our knowledge, such studies have not attracted sufficient attention from researchers working on users' perspectives on cyber security and privacy. In this section, we introduce some selected works we consider closely related non-expert users perspectives on cyber security and privacy, and how these have been analysed in OSN platforms such as Twitter.

Different aspects of non-expert users' cyber security behaviors have been studied by researchers, e.g., conscious/unconscious decisions of non-expert users to recognize different types of cyber threats (Kang et al., 2015; Wu and Zappala, 2018), ignoring necessary updates of important software (Ion et al., 2015), preferring not to use encryption (Wu and Zappala, 2018), hesitance about the use of MFA (multi-factor authentication) (Das et al., 2020), password managers (Alodhyani et al., 2020) and difficulties in configuring security settings such as for biometric authentication Wolf et al. (2019). Jones et al. (2021) warned that insecure behaviors of non-expert users such as above lead to many security vulnerabilities. Hasegawa et al. (2022) conducted a qualitative analysis of 445 cyber security related posts by non-expert users on a question-and-answer website and found the topics of questions ranged from identification and responses to cyber security incidence to usability of security software. We did not see any past research based on large-scale analysis of online discussions on cyber security and privacy of non-expert users.

However, some researchers leveraged public data such as those on Twitter to study different aspects of user perspectives on cyber security and privacy. For instance, Saura et al. (2021) explored the Twitter-verse for cyber security issues commonly discussed by home users in the context of smart living environments. Kowalczuk (2018) used a mixed method to analyze tweets and Amazon reviews of smart speakers users and found enjoyment as the primary reason for using smart speakers. Sriram et al. (2021) conducted a cross-sectional analysis on both Twitter and Reddit data to identify cyber security and privacy con-cerns of end users of IoT devices. Zubiaga et al. (2018) conducted a longitudinal study of tweets between 2009 and 2016 with topic and sentiment analysis to understand the general public's perception of IoT devices. Such research often focused specifically on IoT and smart home devices, so did not cover the wide range of topics including devices that often feature in user discussions.

Some other researchers used public data to monitor and analyze citizens' feelings about security, which covers more physical aspects such as crime in a neighborhood or a city. For instance, Chaparro et al. (2020) and Camargo et al. (2016) explored Twitter data to understand behaviors of online users from specific geo-locations. Greco and Polli (2021) proposed some methods for calculating real-time public perception of cyber security measurement. Such work does not have a smart home focus.

Some researchers focused on privacy-related behaviors of online users, especially about unintentionally leaking personal information on OSN platforms such as Twitter. For instance, privacy features were explored by Caliskan Islam et al. (2014) to detect users' online behaviors and Aghasian et al. (2020) to develop a privacy scoring model for measuring such behaviors. Khazaei et al. (2018) reported how online users who would like to remain private online can be detected, so businesses can respect their privacy preferences. In addition, some researchers also studied leakage of personal information via online users' discussions, e.g., Sharma et al. (2022) (Sharma et al., 2022) reported that Twitter users often unintentionally disclosed personal information online while discussing the COVID-19 pandemic, as observed in four countries (Australia, India, the UK, and the US) and in three different periods of the pandemic (before, during and after the lockdown) in each country.

Another related research topic is machine learning based automatic detection of cyber security related discussions on OSN platforms such as Twitter Alves et al. (2021), Dionísio et al. (2019), Mittal et al. (2016). Most of these types of work are often OSINT (open source intelligence) related and have normally a strong technical focus, instead of users' perspectives.

As reviewed above, although some researchers have leveraged data collected from OSN platforms to study cyber security and privacy-related behaviors of online users, their focus is relatively narrow and did not look at a wide range of topics online users discussed over time. This paper aims to fill such a gap.

## 3. Methodology

As mentioned in Section 1, we decided to focus on online discussions of non-expert users to better capture their perspectives on cyber security and privacy. This required us to develop two classifiers – one 'CySecPriv' classifier for detecting tweets related to the topic (cyber security and privacy) and one 'NonExpertUser" classifier for detecting the target user group (non-expert users), in order to address RQ1. After developing the two classifiers and testing their performance, we applied them to a large set of tweets collected using the Twitter API to produce the dataset we worked with. Topical modeling, sentiment analysis, and additional trend analysis were then conducted on the dataset to answer RQs 2–5.

### 3.1. Classifier selection

For each of the two classification tasks, we trained and tested five candidate classifiers using a labeled dataset in order to identify the best classifier: 1) four traditional machine learning algorithms, including logistic regression, support vector machines (SVM) with a radial basis function (RBF) kernel, random forest and XGBoost, working with n-gram features, and 2) a BERT-based classifier that can extract features via a more automated process.

The four traditional classifiers based on n-gram features were trained, 5-fold cross-validated, and tested using Scikit-learn sckit learn, a popular open-source machine learning package. For the BERT-based classifier, we chose the 'BERT-base' model with 110 million parameters, 12 transformer layers, and 12 self-attention heads (Devlin et al., 2018), implemented as part of the widely used library Huggingface Face. We used Google Colab's NVIDIA Tesla P100-PCIE-16 GB GPU's to run all classifiers.

According to the results shown in Table 1, for both types of classifiers ('CySecPriv' and 'NonExpertUser'), the BERT-based algorithms produced better results. Hence for the later steps we decided to use the BERT-based classifiers to automatically process our dataset.

### 3.2. Data collection

We collected tweets using Twitter's Academic API v2[4]. Due to the rate limits of the API[5], collecting tweets in all 36 months in 2019–2021 could take a long time, likely 36 weeks (more than 8 months) according to our experiments. Therefore, we decided to collect tweets over a two-month period each year. In order to determine which two representative months to select, we considered two factors. Firstly, according to some recent statistics (Sabanoglu; Salesforce; Ward, 2022), the level of product sales is often the highest in November and December, i.e., at the end of the year just before the Christmas holiday period. If we assume that users would spend some time using new devices during the holiday period, they would more likely to share their experiences and to ask questions on social media in January and February. Secondly, since we would like to collect data that can support a comparison of online discussions of English-speaking users before and during the COVID-19 pandemic, we looked at the time when the global lockdown started worldwide. This seems to be in March 2020 for most countries where English-speaking users lived (BBC, 2020). Based on the two factors, we decided to choose January and February as the two-month period, which allowed us to have two years before the first wave of global COVID-19 lockdowns (2019 and 2020) and one year during the pandemic (2021).

In order to retrieve relevant historical tweets in the past, we needed to use Twitter's Search API, which required us to use some keywords as search terms. Determining the most appropriate search keywords was not trivial, as non-expert users do not use technical terms expert users would use for discussing cyber security and privacy-related topics. Hence, we decided to use some general computing keywords and names of different computing devices non-expert users often use at home. Rather than defining such keywords ourselves, we used the following sources as the basis to derive such keywords, in order to avoid any biases we may have.

1. *Some keywords selected from an empirical study*: We conducted an online anonymous survey with 50 participants using the Prolific survey platform[6], on frequently used keywords by non-expert users while searching for cyber security and privacy-related queries online. This survey was approved by the University of Kent's Central Research Ethics Advisory Group. Out of 144 keywords reported by the participants, we selected 23 ones based on their nature (computing/security-related term), and the frequency of use (used by more than one participant). We avoided using too broad keywords such as 'Google', 'Internet' and 'Windows'.

---

[4] https://www.developer.twitter.com/en/products/twitter-api/academic-research
[5] There were multiple limits for our approved Twitter Academic API account: 1 app per account, 10 million tweets/month per account, 300 requests / 15 minutes per app, 1 request/second per app, and 500 results per response.
[6] https://www.prolific.co/

**Table 1**
Performance comparison of different machine learning models for the two classification tasks of our work

| Model | Precision | Recall | F1-Score | Model | Precision | Recall | F1-Score |
|---|---|---|---|---|---|---|---|
| **BERT-based** | 0.92 | **0.92** | **0.91** | **BERT-based** | **0.94** | 0.95 | **0.94** |
| Logistic Regression | 0.90 | 0.84 | 0.87 | Logistic Regression | 0.83 | 0.90 | 0.86 |
| SVM (RBF kernel) | 0.92 | 0.76 | 0.83 | SVM (RBF kernel) | 0.68 | **1.0** | 0.81 |
| Random Forest | **0.94** | 0.61 | 0.74 | Random Forest | 0.70 | **1.0** | 0.83 |
| XGBoost | 0.90 | 0.86 | 0.88 | XGBoost | 0.73 | 0.98 | 0.84 |
| (a) 'CySecPriv' classifier | | | | (a) 'NonExpertUser' classifier | | | |

2. *A number of lists of most used computing devices at homes reported in the research literature*, including:
   - a list of different types of smart home devices shown in Table 3 of (Huang et al., 2020a),
   - a list of commonly used smart devices shown in Table I of (Gai et al., 2018),
   - a list of smart devices used at home in multi-user scenarios, as shown in Table I of (Huang et al., 2020b),
   - a list of smart devices shown in Fig. 3 of (Cannizzaro et al., 2020), and
   - a list of common devices owned by smart home users shown in Table 1 of (Zheng et al., 2018).

3. *Commonly used computing devices included in the following reports from Statista*:
   - a 2020 report on market shares of electronic devices used by internet users in the UK to go online (O'Dea, 2021),
   - a 2020 report on the average number of devices residents have access to in UK households (Laricchia, 2022),
   - a 2020 report on devices used to access Wi-Fi in UK homes (Alsop, 2021), and
   - a 2021 report on devices used by global users to access internet (Johnson, 2022a).

Based on the above sources and our general knowledge on relevant keywords and computing devices used by non-expert home users, we derived 37 unique keywords: 'account', 'Amazon Alexa', 'Amazon Echo', 'app', 'breach', 'e-reader', 'Google Home', 'iPad', 'iPhone', 'android', 'laptop', 'smart mobile', 'password', 'smart TV', 'tablet', 'WiFi', 'Bluetooth', 'smart camera', 'smart watch', 'smart doorbell', 'smart switch', 'Chromecast', 'smart speaker', 'smart hub', 'smart light', 'printer', 'smart thermostat', 'VPN', 'smart kettle', 'smart refrigerator', 'smart washing machine', 'smart meter', 'smart toy', 'smart door lock', 'smart baby monitor','smart plug', and 'games console'.

Using the 37 keywords with Twitter's Academic Search API, we obtained 13.7 million tweets for further processing. We decided not to collect retweets as they include only repeated discussions. In other words, our collected data consists of original tweets, replies and quoted tweets (without the quoted original tweet), and other meta-data such as 'authorId', 'userName', 'createdAt', 'location', 'publicMetrics'.

### 3.3. Data cleaning, tokenization and vectorization

To facilitate feature extraction for the candidate classifiers, we needed to conduct some pre-processing steps, including data cleaning, tokenization and vectorization. Special characters, hyperlinks, references to audio and video files, and tweets without any text but only hashtags and/or mentions were taken out as the first step of data cleaning. Emojis were transferred to their equivalent words or phrases in English using the 'emot' library (Shah, 2022).

For the n-gram-based classifiers, we also applied stop-word removal and lemmatization to prepare them for tokenization. The processed data was then tokenized using the BOW (bag of word) method (Sethy and Ramabhadran, 2008) and vectorized using the TF-IDF (Term Frequency and Inverse Document Frequency) vectorizer of the Scikit-learn library[7] to prepare a matrix of TF-IDF features. This TF-IDF matrix is then fed into the four n-gram based classifiers as input features.

The pre-processing of data differed for the BERT-based classifier since it works in a very different way from the traditional machine learning algorithms we used. BERT, a neural network-based encoder technique reported in Devlin et al., 2018, works with a bidirectional contextual word embedding modeling method. The model essentially stacks a series of encoder structures based on the transformer architecture (Vaswani et al., 2017) and pre-trains the model by masking out a certain percentage of the words, which forces the model to learn the words. Deliberate retention of special characters such as question marks, exclamation marks, and commas were implemented and no stop-word removal techniques were used to maintain the BERT model's original effectiveness (Hofstätter et al., 2020) as the model inherently calculates its algorithm by using context-based attention. Once the pre-trained model was loaded, it was fine-tuned with the help of our context-specific pre-processed tweets. This was then fed to the Hugging Face transformer library to implement TensorFlow's BERT-based sequence classification module (Google, 2022) to classify our data.

After completion of the pre-processing steps, we noticed many duplicate tweets in our data, which were possibly bot-generated. We decided to remove the duplicate tweets before any further processing. Finally, after the pre-processing and deduplication procedure, we obtained 13.7 million tweets as the raw data for labeling and classification to produce the dataset of cyber security and privacy-related tweets.

### 3.4. Developing classifiers: BERT-based

Following common practices Devlin et al. (2018), Hao et al. (2019), we decided to fine-tune the pre-trained BERT model using our cyber security-specific dataset. We used the deep learning library TensorFlow Keras[8] and ktrain[9] (a lightweight wrapper for Tensorflow Keras) to build our classifiers. We did not freeze any of the existing pre-trained layers and hence the weights of all layers were updated during the backpropagation. Our processed tweet dataset was applied to the pre-trained BERT module for fine-tuning to get to the final classifiers (see Fig. 1).

The learner object in the ktrain library used the training and validation data to fine-tune the classifier with a small batch size of 16. The other important parameters for fine-tuning are mentioned below.

**Learning rate:** This is an important parameter that controls the continually updated error proportion in the model weight at the end of each batch of the training examples. A large learning rate gives a faster running model, but might produce sub-optimal model weights. Conversely, a smaller learning rate might produce a more optimal set of weights but could be very slow to run.

---

[7] https://www.scikit-learn.org/stable/modules/generated/sklearn.feature_extraction.text.TfidfVectorizer.html
[8] https://www.tensorflow.org/text/tutorials/classify_text_with_bert
[9] https://www.github.com/amaiya/ktrain

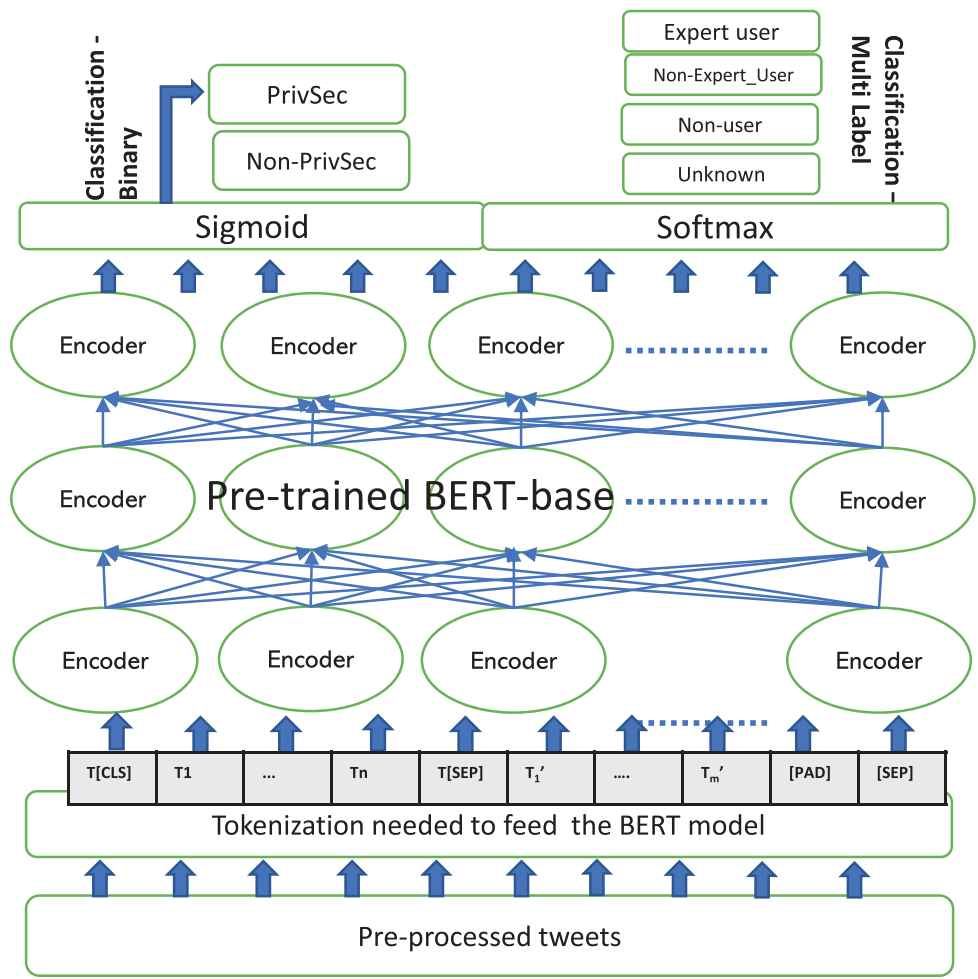

**Fig. 1.** A diagram showing how the 'CySecPriv' and 'NonExpertUser' classifiers are constructed.

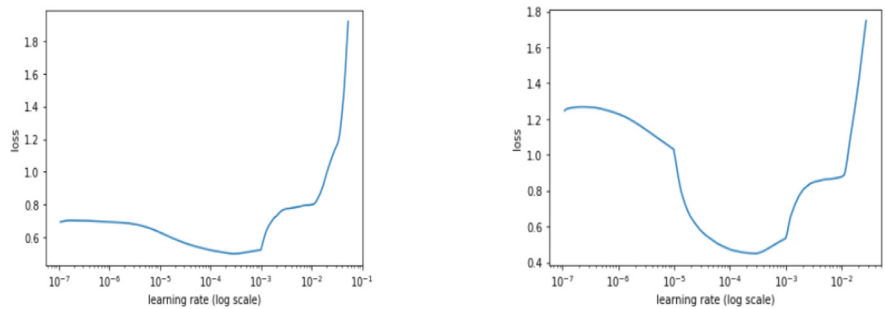

**Fig. 2.** Visually inspect loss plot and select learning rate associated with falling loss.

The '*learner rate find*' module facilitated finding a balanced learning rate of '5e-5' (see Fig. 2) for both the 'CySecPriv' classifier and 'NonExpertUser' classifier.

**epoch number:** We experimented on different epoch numbers (1–20) and noticed the best output was from 2 epochs where there was the least amount of gap between the loss in the training and validation dataset indicating a good fit in both the classifiers case.

### 3.4.1. Cyber security and privacy ('cysecPriv') classifier

To develop a classifier, we first need a labeled dataset. Based on our expert knowledge, we decided to use a number of criteria to manually label a tweet as 'CySecPriv' or 'NonCySecPriv', depending on, if the tweet fulfills any of the following points:

1. at least one term from the following established lists of cyber security and privacy-related terms:
   - https://www.ncsc.gov.uk/information/ncsc-glossary
   - https://www.bsigroup.com/en-GB/Cyber-Security/Glossary-of-cyber-security-terms/
   - https://www.getsafeonline.org/glossary/
   - https://www.niccs.cisa.gov/about-niccs/cybersecurity-glossary
2. at least one keyword included in the list of terms of the cyber security taxonomy reported in (Mahaini et al., 2019)[10],
3. at least one hashtag related to any of the above terms, and

---

[10] https://www.cyber.kent.ac.uk/research/cyber_taxonomy/

4. at least one phrase or sentence that clearly describes a cyber security or privacy related topic, i.e., a tweet like 'my computer must have been infected after I downloaded a movie'[11].

The labeling process was completed in two different stages. First, 9,000 tweets (1,500 from each of the six months in our dataset) were randomly selected and labeled by the first author according to the above-mentioned criteria. This led to an imbalanced dataset with only 5% of tweets being labeled 'CySecPriv'. To produce a more balanced dataset for training and testing purposes, we collected a set of 18,000 (3,000 from each month) tweets based on a different set of cyber security related keywords (within the scope of the first two items of the list 1) and labeled them separately. Some tweets (676) in this set were discarded because either they were duplicate tweets or they did not contain any real original content excepting '@mentions' and/or hashtags. Labeled data from both stages were combined to produce our final labeled dataset with 26,324 tweets, including 14,041 CySecPriv tweets and 12,282 NonCySecPriv tweets. The database is still unbalanced, but not significantly so. Our experiments showed that such a dataset already worked relatively well, so we did not apply any over- or under-sampling to make the dataset perfectly balanced.

### 3.4.2. 'NonExpertUser' Classifier

For this classifier, we followed a multi-class approach to label a Twitter account into four different classes based on its meta-data (username, display name, user profile and location), as shown below.

1. Non-expert user: a user account satisfying both of the following criteria:
   (a) the profile is a personal account, indicated by the use of a first-person pronoun (i.e., 'I', 'me') at least once, or a noun or a phase representing a person, e.g., a name like 'Bob' or a relation such as 'wife' or 'son'.
   (b) the profile clearly suggest that the account owner is not an expert on cyber security or privacy, e.g., if the profession is declared to be 'actor', 'painter', 'musician' or 'trader'. (There might of course be some cases where the candidate does not declare themselves to be a security expert, but actually are in the offline world.)
2. Expert: a personal account (following the above criterion 1(a) for non-expert users) that used at least one recognized cyber security or privacy-related term (as described in the keyword list 1 of the 'CySecPriv' classifier) to describe their work, education or expertise, e.g., 'security expert', 'cryptographer', 'hacker' or a 'student studying a cyber security course'.
3. Non-Person: a user account that belongs to an organization, a group of people, or a community.
4. Unknown: when the information in the user profile is insufficient to draw any conclusion about the type of account.

To help facilitate the manual labeling process, we leveraged the widely used NER (named entity recognition) module in the Spacy library GmbH to automatically detect entities related to persons and organizations.

To support the training and testing of the non-expert user classifier, we sampled 10,200 accounts randomly from the raw dataset and collected their meta-data. The meta-data was pre-processed to exclude non-ASCII characters and special characters that are not exclamation marks, question marks, full stop signs, double and single quotation marks, and to convert emojis into equivalent English phrases using the 'emot' library. The pre-processed meta-data was then input to the NER module of SpaCy to detect any named entities. Taking the NER tags as useful references, the sample accounts were then manually labeled by the first author into the above-mentioned four classes, leading to 7,860 labeled as non-expert users, 825 as experts, 1,143 as non-person, and 394 as unknown. The labeled dataset was highly imbalanced, but we decided not to modify (over- or under-sample) it as current research has shown evidence (Aduragba et al., 2020; D'Sa et al., 2020; Madabushi et al., 2020; Oak et al., 2019) that some machine learning models such as BERT-based ones can still work well with imbalanced data. In addition, although being a multi-class classifier, for our purpose it was used more like a binary classifier for detecting one class (non-expert users). For such a binary classifier, the dataset imbalance is less serious: 7,860 vs. $825 + 1,143 + 394 = 2,362$.

The labeled data was then fed as input to different candidate classifiers to identify the best performing algorithm. All five classifiers did reasonably well on the imbalanced dataset, although we found that the BERT-based classifier performed much better than the four traditional classifiers, as shown in Table 1. More detailed explanation on the performance metrics of the classifiers are given in Section 4.

### 3.5. Applying classifiers

After obtaining the trained and tested 'CySecPriv' and 'NonExpertUser' classifiers, we applied them sequentially to the pre-processed dataset of 13.7 million tweets to produce the actual dataset for further analysis. The final dataset consists of 435,207 (3.17% of the total tweet dataset) CySecPriv tweets written by 337,604 non-expert users. Fig. 3 demonstrates the number of tweets after each step of the data curation and the number of authors after applying the classifiers.

Due to the non-perfect accuracy of the classifiers, the dataset must also include a small portion of false positive samples and have missed some false negative samples. Taking this into consideration, later we will avoid results that can be affected by such false positives, e.g., for topical modeling results, any topics with a small number of tweets may be more affected by false positives so become less reliable.

### 3.6. Topic modeling

Once we had the final dataset, we investigated important topical areas that attracted non-expert users' attention (RQ2 and RQ3). We tried two popular topic modeling algorithms for this purpose: latent Dirichlet allocation (LDA) (Blei et al., 2003) and non-negative matrix factorization (NMF) (Kuang et al., 2015), both having been widely used and NMF being reported to perform better in some tasks especially for processing shorter texts like tweets (Chen et al., 2019). We found that LDA provided semantically more interpretable results compare to NMF, while NMF was generally faster in processing the outputs. As the interpretability of the results is more important than speed for our purposes, we decided to use LDA as the topical modeling method. LDA requires the number of topics as an input parameter, so we needed a way to determine the best value of the number. We used Gensim[12], a popular topic coherence model tool, to determine a relevant *k* while processing all topic modeling cases. In addition, an R package pyLDAvis (Sievert and Shirley, 2014) was used to visualize and interpret the results of LDA.

The LDA algorithm was used to analyze topics in our final datasets in three different ways: (1) tweets in all three years together to get the overall picture, (2) tweets in each of the three different years separately to allow yearly trend analysis, and (3) 2019 and 2020 tweets combined vs the 2021 data to study any changes of topics before the global COVID-19 lockdowns and during the pandemic.

---

[11] Note that this quoted tweet is an artificially created illustrative example to avoid exposing any real tweet in our dataset for data protection reasons.

[12] https://www.radimrehurek.com/gensim/

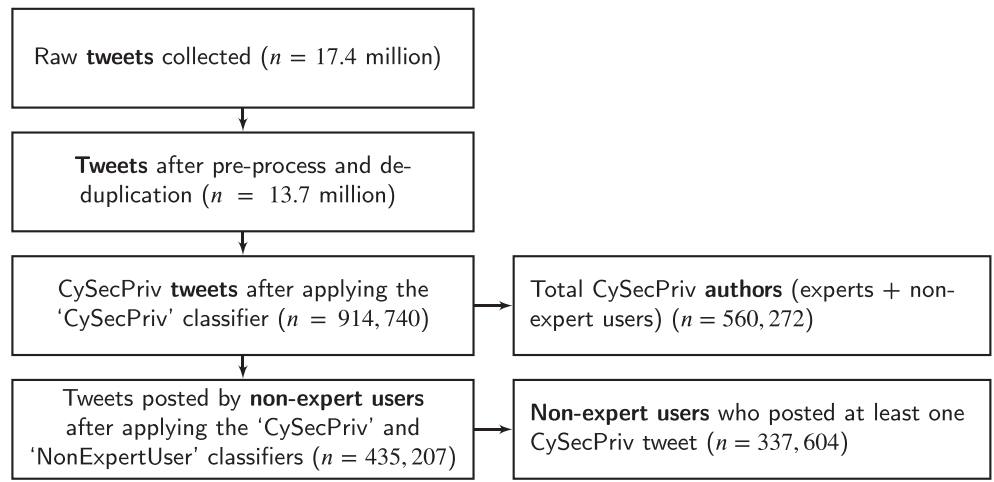

**Fig. 3.** The number of tweets after each step of the data curation process and the number of corresponding authors for the last two steps.

### 3.7. Sentiment analysis

To address RQ4 and RQ5, we conducted sentiment analysis on the final dataset. We experimented with two sentiment analysis algorithms, VADER (Valence Aware Dictionary for Sentiment Reasoning) analysis (Hutto and Gilbert, 2014) that is a popular lexicon-based sentiment analysis library for Twitter-like datasets and 'bert-base-multilingual-uncased-sentiment' Town, a BERT-based sentiment analysis algorithm implemented by the NLP Town[13] for the Hugging Face library. VADER classifies an input text into three different classes, i.e., 'positive', 'negative' and 'neutral', whereas the BERT-based algorithm produces a five-leveled sentiment score (between -2 and 2), with -2 being totally negative and 2 totally positive[14]. Results produced by the two algorithms are very different, so we decided to evaluate their performance against a manually constructed benchmark.

To decide which sentiment analysis algorithm to use, we conducted a comparison between them using some ground truth labels for tweets with 'positive', 'negative' and 'neutral' sentiment. Such labels were produced by the first author of the paper by manually inspecting 500 randomly selected samples, which were classified as totally positive (2), totally negative (-2) or neutral (0) by the BERT-based algorithm. The ground truth labels were compared with the outputs from both sentiment analysis algorithms to calculate an agreement rate, i.e., the percentage of the ground truth labels match the outputs of a given algorithm, as an accuracy metric. The results revealed that the BERT-based model achieved an agreement rate of 71%, 10% higher than VADER that achieved an agreement score of 61%. The results are aligned with results from past studies conducted by other researchers (Crocamo et al., 2021; Nemes and Kiss, 2021). Therefore, we decided to use the BERT-based algorithm for our sentiment analysis experiments.

### 3.8. Qualitative analysis

In order to gather more insights about some of the important topics we observed from our earlier quantitative analysis we decided to conduct a qualitative analysis. Our methodology follows the four-phased approach proposed in (Andreotta et al., 2019). We decided to qualitatively inspect randomly selected tweets belongings to six LDA topics, including 1) four cyber security-related topics that were frequently discussed in all three years – 'VPNuse', 'DeviceAccSec' and 'WifiPass', 'HomePrivSec' and 2) two topics that appeared only in pre-Covid period (2019-20) or during pandemic period in 2021 – 'External Stakeholders' 2019/2020 for 2019-20 and 'HelpRelated' for 2021, as presented in Fig. 6. We randomly extracted 500 tweets from each of the 5 selected topics. This led to in total $500 \times 3 \times 4 + 500 \times 2 = 7,000$ tweets for qualitative analysis. For the analysis itself, we followed the thematic analysis method proposed in (Braun et al., 2019). Important keywords from each topic were highlighted and assigned to different codes first. For example, a tweet on forgetting password was assigned to a code 'password' and a tweet on sharing a password with a family member to 'sharing behavior'. Eventually, these codes were compared and grouped into a number of coherent themes. For example, all Wi-Fi related codes such as 'WiFi neighbour/family/friends', 'Wi-Fi awareness', 'Wi-Fi password', 'Wi-Fi Device', 'Wi-Fi Issue/solution/help' were grouped in the theme 'Wi-Fi related security'.

## 4. Results

In this section, we report detailed results of our work, including detailed performance comparison results of the 'CySecPriv' and 'NonExpertUser' classifiers based on different machine learning methods, and quantitative and qualitative analysis results after applying the BERT-based classifiers to our Twitter dataset.

### 4.1. Performance of 'CySecPriv' and 'NonExpertUser' Classifiers (RQ1)

Identifying privacy and security related tweets posted by non-expert users was our first research question. To that extent, we developed a classifier to first determine the cyber security nature of a tweet, and then a second classifier to identify the author as a non-expert user. The following paragraph explains the performance success of our classifiers. According to the results in Table 1a, we can see that the BERT-based classifier performed the best, so we decided to use it for our analysis. This classifier achieved a precision score of 0.90, a recall score of 0.93 and

---

[13] https://www.nlp.town/
[14] The algorithm originally returns a score between 1 and 5. We decided to shift the range by -3, so the range become [-2,2] and a negative/positive value means negative/positive sentiment, which we consider more interpretable.

**Table 2**
The yearly trend of CySecPriv tweets and non-expert (NE) authors of such tweets (2019–2021), showing a substantially increased level in 2021 (during the COVID-19 pandemic) compared with 2019 and 2020 (before the first wave of global COVID-19 lockdowns), in both relevant tweets and non-expert authors involved.

| Year(s) | #(Initial Processed Tweets) | #(CySecPriv Tweets) | #(NE CySecPriv Tweets) (%) |
|---|---|---|---|
| 2019 | 4,690,397 | 274,279 | 90,539 (2.5%) |
| 2020 | 4,439,719 | 284,006 | 110,881 (2.5%) |
| 2021 | 4,589,582 | 356,455 | **233,787** (**5.09%**) |
| 2019-21 | 13,719,698 | 914,740 | 435,207 (3.17%) |

(a) CySecPriv tweets posted by non-expert users

| Year(s) | #(CySecPriv Authors) | #(NE CySecPriv Authors) |
|---|---|---|
| 2019 | 202,917 | 90,539 (45%) |
| 2020 | 186,964 | 87,513 (47%) |
| 2021 | 247,417 | **180,325** (**73%**) |
| 2019-21 | 560,272 | 337,604 (60%) |

(b) The number of non-expert authors

an F1-score of 0.92, a reasonably high score to classify and label our tweets. We compared our results to other similar studies using BERT-based classifiers to evaluate our model. We noticed that, in general, BERT-based models achieved an F1-score between 0.83 and 0.95 (Dionísio et al., 2019; Ghourabi, 2021; Husain and Uzuner, 2021; Kalepalli et al., 2020; Mozafari et al., 2020; Sriram et al., 2021). Similarly good performance results were observed for the 'NonExpertUser' classifier. As shown in Table 1b, the best-performing classifier is also the BERT-based one, which achieved an F1-score of 0.94 when it is used to predict non-expert users (versus the other three classes).

As mentioned in the previous section, after applying the two classifiers to the 13.7 million raw tweets, we obtained 435,207 tweets written by non-expert users. Among all the CySecPriv tweets authored by non-expert users, we noticed a sharp (more than doubled) increase both in the absolute number and the relative percentage of CySecPriv tweets from 2019 and 2020 to 2021, as shown in Table 2. If we look at the proportion of non-expert users in the total CySecPriv authorship, who posted at least one CySecPriv tweet, we can also see a similar increase from 2019 and 2020 to 2021, doubled in the absolute number and increased by 62.2% (vs 2019) and 55.3% (vs 2020) in the relative percentage. The increase during the COVID-19 pandemic (2021) may be attributed to the increased use of digital technologies at home due to the global lockdowns that led to more cyber security and privacy challenges and concerns for non-expert users.

### 4.2. Non-expert users' cyber security and privacy related discussions and trend analysis (RQ2 and RQ3)

Having identified the relevant tweets, we explored the results using topic modeling to understand the types of discussions non-expert users were engaged in on Twitter. We examined the topics discussed both quantitatively and qualitatively. The quantitative analysis involved the entire curated dataset while the qualitative exploration considered a representative subset of data (see Sub-Section 3.8). In the following two sub-subsections, we discuss the results of both investigations, in greater detail.

#### 4.2.1. Quantitative analysis

Our LDA analysis investigated the datasets from the following three different angles:

I analysis of the whole dataset,
II analysis of data over time (yearly), and
III comparison of data in the first two years (January 2019 to February 2022, before the first wave of global COVID-19 lockdowns) and during the pandemic (January and February of 2021).

These separate analyses led to discovery of different patterns, e.g., the third analysis let us see how 'Help-related' tweets became a major topic of non-expert users' online discussion on Twitter in 2021, much more so than before the COVID-19 pandemic. Similarly, the second analysis highlighted the trend in certain topics such as the continuous increasing use of 'VPN-related' queries over time, which may not be explained by the COVID-19 pandemic alone. Results of all these analyses are discussed with greater details below and shown in Figs. 4, 5 and 6.

*(I) Analysis of the whole dataset:*

The results depicted in Fig. 4 demonstrate common discussion areas of non-expert Twitter users pertaining to cyber security and privacy. Interestingly and unexpectedly, the top topic turned out to be the use of VPNs ('VPNuse', 21%), which is followed by device account security ('AcctPrivSec', 13%), laptop security ('LaptopSec', 10%), Wi-Fi and password security (9.2%), and cyber security and privacy matters at home ('HomePrivSec', 8.6%). In addition, cyber security and privacy matters related to different computing devices is also an important cluster of topics: iPhone ('iPhoneSec', 8%), general devices ('DevicePrivSec', 7.3%), Android ('AndroidPrivSec', 6.3%) and mobile apps ('AppPrivSec', 6%). Considering the second most used topic 'AcctPrivSec' is also device-related, it may be the case that device-related discussions are even more common than those related to the top-most topic 'VPNuse'. Note that the boundary between different topics is not clear-cut, so there are often overlaps between different discussions, e.g., a tweet in our dataset about how to use an IP camera and access details through VPN could come either under 'DevicePrivSec' or 'VPNUse'.

*(II) Analysis of data over the three years:*

Fig. 5 gives a compact view of the top topics calculated for each year along with the percentage of tweets in each topic. As in the case of the overall results, the top-most topic for all the three years remains 'VPNuse'. Inspecting the overlaps between different topics quantitatively and qualitatively (see Section 4.2.2 for the latter), we observed that the 'VPNuse' topic overlaps heavily with other topics, indicating VPNs were used in many different contexts. For example, 'VPN' as a keyword was mentioned in 35% of tweets on the 'HelpRelated' topic in 2021 and 16% of the total tweets of the 'Wifi-Pass' topic in 2020. While the topic distributions across the three individual years look similar, there are some noticeable differences. For instance, the percentage of the 'WifiPass' topic decreased from 10.4% in 2019 to 9.5% in 2021, substituted with broader discussions on new topics such as 'HomePrivSec' (3.6%), 'InternetSec' (2.2%) and 'HelpRelated' (2.7%).

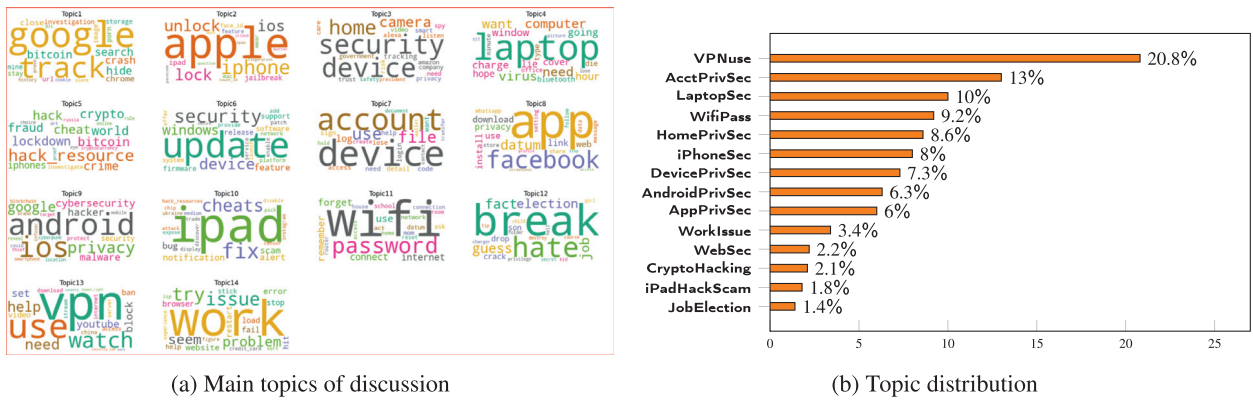

(a) Main topics of discussion    (b) Topic distribution

**Fig. 4.** Topics derived from the full dataset (2019-21).

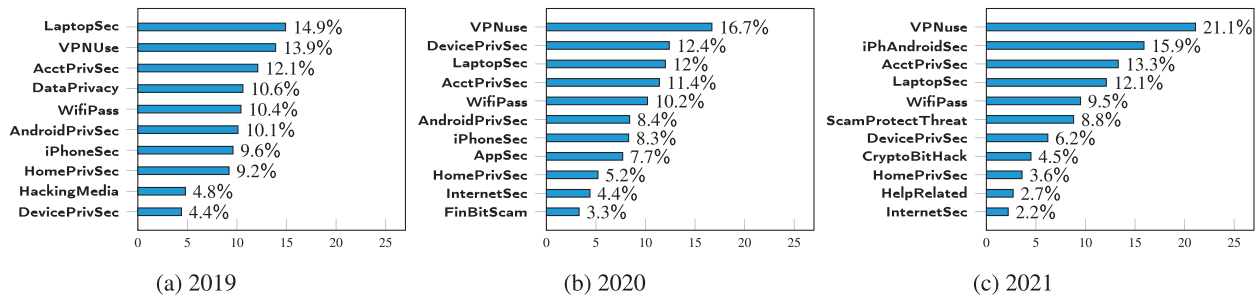

(a) 2019    (b) 2020    (c) 2021

**Fig. 5.** Topic distribution for each of the three individual years (2019-2021).

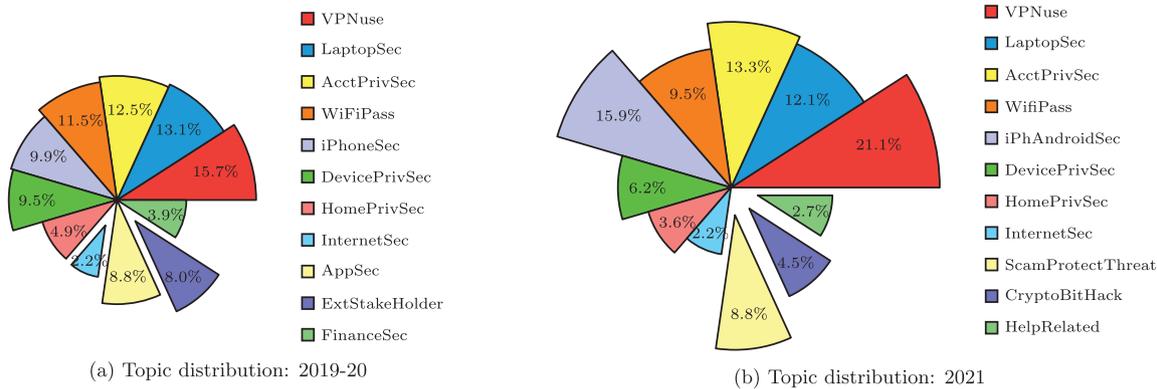

(a) Topic distribution: 2019-20    (b) Topic distribution: 2021

**Fig. 6.** Topic distributions in 2019-20 (before the first wave of global COVID-19 lockdowns) and in 2021 (during the COVID-19 pandemic).

*(III) Comparison of tweets before and during the COVID-19 pandemic:*

The topic modeling results presented in Fig. 6 show statistics of topics before the first wave of global COVID-19 lockdowns (2019-20) and during the pandemic (2021). It is evident from the figure that, several topics appeared to be common in both periods, e.g., 'VPNuse', 'LaptopSec', 'AccPrivSec', 'iPhoneSec', and 'WifiPass', indicating that the nature of the frequently discussed topics before and during COVID period have not changed much. However, we can also notice that, there are topics which are different, e.g., a number of new topic emerged during the pandemic period, including 'HelpRelated' (2.7%) containing tweets about peoples' query, help or advice related tweets, and 'ScamProtectThreat' (8.8%) relating to different types of scams and security threat people encountered and ways to protect themselves. These may be attributed to more cyber security and privacy issues encountered by non-expert users due to an increased level of digital activities during the COVID-19 pandemic. There is a notable increase in the topic area of 'VPNuse' during the 2021 pandemic period, which suggests that non-expert users' use of VPN may have increased during this period. Another major trend emerged is that non-expert users were concerned by and engaged with discussions of a wide range of topics, such as cyber security and privacy issues of both traditional computing devices and 'smarter' devices, networking behaviors and authentication techniques such as passwords, privacy enhancing techniques and tools such as VPNs, seeking help and support on different privacy and cyber security related topics, Wi-Fi-related topics, financial security involving external stakeholders, cyber security specific to home, mobile device security and web browser security. Evidences of these discussions could be found in topics across the years.

#### 4.2.2. Qualitative analysis

As mentioned in Section 3, we used thematic analysis to code the selected six topics for our qualitative analysis, which resulted in seven different themes and provided further insights into the topic categories. These details are discussed below. Although we collected topics generated by LDA, after manually going through the data we observed several overlaps in the topics. For example, we had 500 records on help-related topics but after the thematic analysis we found 1,852 tweets that could be assigned to the theme of 'Help and support needed' and otherwise were in either 'AcctPrivSec', 'VPNuse', 'ExtStakeHolder' or 'WifiPass' in the LDA results. Such overlap were envisaged due to the automatic nature of analysis done in LDA topic modeling.

*(I) Help and Support Needed:* A number of queries and discussions (222 tweets, 12% of 1852 help-related ones) on laptops, desktops, tablets and routers involved cyber security problems about passwords and user authentication, where non-expert users expressed their inability to cope with the requirements of multiple password changes and the use of OTPs (one-time passwords). Users sought help on virus-related problems, and their inability to distinguish between scams and genuine requests from legitimate parties sometime indicating people's general awareness of scam messages, but also their lack of technical skills to recognize scams and consequently help themselves. There were a number of discussion (333 tweets, 18% of 1,852) involving questions and advice related to Wi-Fi security especially about secure Wi-Fi in public places from a user perspective, and also seeking help on forgotten passwords, broken Wi-Fi security, how to stop neighbors from breaking into Wi-Fi, handling security notifications, etc. Cyber security related questions regarding web browsers on PCs were another major area of discussions, often about seeking advice on how to clear caches and cookies, how to deal with adware and how to use private browsing. VPN-related questions covered a big portion (482 tweets, 26% of 1,852) involving questions on downloading and setting up VPNs, different operational problems, seeking help on what type of VPNs to install, etc.

*(II) Awareness on Password Protection:* User authentication, and passwords particularly, have been prominent topics of cyber security related discussion for a long time (Zimmermann and Gerber, 2020). 11% of the total data we analyzed (826 of 7,000) were password-related discussions. For the tweets we analyzed in this study, we noticed that only a very small number of tweets (32, 3.9% of 826) referred to modern authentication schemes based on biometrics or multi-factor authentication (MFA). Wi-Fi security is a major topic (490 tweets, 60% of 826), where non-expert users expressed the risks associated with leaked Wi-Fi passwords and their worry or fear of such passwords being compromised. Password managers (PMs) appeared to be a popular sub-topic as well. Tweets on PMs covered both positive and negative aspects, covering areas such as browser-based PMs or dedicated software tools, advice on the best PMs, open-source and commercial PMs, selective use of PMs, and how to use PMs. A very small number of tweets (16, 1.9% of 826) advised using PMs as a good cyber security habit or voiced doubts on their usefulness. Tweets involving general passwords were mainly about how users were frustrated to remember difficult passwords, unable to cope with managing passwords for multiple devices or expressed distrust of stakeholders who manage PMs. In line with the discussions in some recent studies (Zimmermann and Gerber, 2020), we noticed that, non-expert users in general were not enthusiastic about more modern authentication methods such as MFA, which may be attributed to the substantial efforts needed for setting things up, compared to the simplicity of using more traditional knowledge-based methods such as textual passwords.

*(III) Wi-Fi related security:* 1852 (26% of 7,000) tweets asked for help and advice on locating Wi-Fi access points, (re)configuring Wi-Fi settings, connecting and disconnecting different devices to and from Wi-Fi access points, and cyber security issues of Wi-Fi routers (e.g., compromised routers). Some tweets under this theme discussed ways of hacking into neighbors' Wi-Fi (e.g., for using neighbors' Wi-Fi illegally to watch paid TV programs) and how to protect themselves from this kind of attacks. Another aspect discussed under this topic is the behavior of some non-expert users in sharing their Wi-Fi passwords with trusted people such as guests, friends and roommates, without any heed to potential security problems. Granting important user credentials to trusted people may seem harmless, and it is one of the usual way of sharing Wi-Fi access points at home, but it can potentially open up possibilities of insecure storage, extension of admin privileges (unless a separate guest account is used). Some interesting tweets in our set talked about how people took photographs of wi-Fi credentials wherever they visited and their phone was full of these details from friends and families. Overall, it seemed that while some non-expert users had some understanding of cyber security issues involving the use of Wi-Fi, many of them had either a low level of awareness or were unsure of how to protect themselves from potential cyber security problems. Setting up an auto-connect, time-limited guest account could solve this problem but unfortunately not all Wi-Fi provider offers this choice and if on offer, they are quite complicated to follow up.

*(IV) Privacy:* Privacy as a topic was discussed by a number of (2,206, 32% of 7,000) non-expert users. Privacy concerns around smart home devices, such as Amazon Echo, Google Nest, smart locks, smartwatches, smartphones and smart TVs, are one of the main reasons behind privacy-related discussions, with a majority of (1,876, 85% of 2,206) all tweets were attributed to discussions related to these areas. Users expressed their mistrust, frustration and awareness or unawareness about data collection, storage and transfer to third parties from these devices. Keywords such as 'scared', 'panic', 'trust' and 'annoy' made a regular appearance within these tweets. On a positive side, users also discussed or advised others on how to improve privacy of these devices. A number of (318, 14% of 2,206) tweets were related to possible data breaches, trust in data collectors, data leaking, data theft, secure data storage and transfer. Webcams and laptop privacy also caught many users' attention, with some of them discussing about privacy filters for laptops and webcam privacy covers. Several tweets mentioned or asked about how VPNs could help enhance privacy and whether it is worth using it for that purpose.

*(V) External stakeholders:* This was an interesting topic covering a wide range of discussions. Such tweets referred to different stakeholders who are in some way responsible for cyber security and privacy of users. The stakeholders mentioned include governments, companies and other third parties who were not part of the non-expert users' operational environments but impacted the way their cyber security and privacy related issues were generated and handled. The activities and impacts of governmental bodies are one of the frequently discussed sub-topics. These include governmental apps snooping on phones and devices, diminishing trust of users on governments handling of their personal data, outdated security certificates on governmental websites, and also low-quality information on governmental websites such as mis-spelling 'Smart meters' as 'Spy meters'. Many users were very vocal about companies not caring about their privacy and data security, tracking user devices to collect personal data covertly, and even pre-installing spyware on user devices.

*(VI) Smartphones and other smart devices:* A lot of discussions (1,876, 26% of 7,000) in our samples were related to iPhone and Android smartphones and other smart devices. Many discussions focused on Android's lack of security (e.g., such devices being vulnerable to malware or spyware), and ways to protect against such security problems. Although many posts were positive about iPhone security, others placed their mistrust in a growing level of security mishaps and vulnerabilities in iPhone. Location tracking using the phone was also discussed sometimes as a good feature for locating lost phones but was also labeled as a harmful feature. Smart speakers such as Amazon Alexa are one of the frequently mentioned smart devices under this topic, and tweets were on dis-cussions of data privacy, security updates, third party involvement in data sharing, and following appropriate security behaviors while using smart speakers.

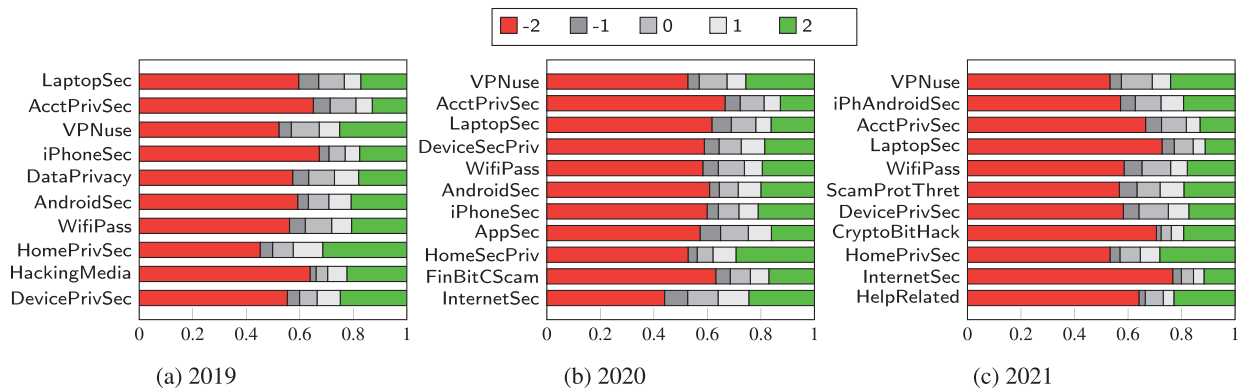

**Fig. 7.** Sentiment analysis results for each of the three years (2019-21).

*(VII) VPN related discussion:* As mentioned in Section 4.2.1, VPN was the most discussed topic as shown in the quantitative analysis, so it was not surprising that we encountered many tweets (2,006, 29% of 7,000) under this topic in our qualitative analysis. Most discussions on VPNs revolved around people trying to get help to download, install, or configure VPNs at home. Various operational issues and general advice on type of VPNs to use or purchase, need of VPNs in different scenarios and the use of VPNs for watching and streaming different programs to avoid privacy issues were also quite popular. The use of VPNs to avoid or override local restrictions on specific data or resource use was a popular topic, too. Many of these tweets (489, 24% of 2,006) advised on the what, how and why's of VPNs, responded to work-related queries, and some also expressed distrust of VPNs.

### 4.3. Sentiment analysis of non-expert users' discussions on cyber security and privacy and trend analysis (RQ4 and RQ5)

Having identified the topical areas non-expert users were interested, we investigated their sentiment with relation to these topics. As mentioned in Section 3, we used an off-the-shelf BERT-based sentiment analysis classifier for this purpose. We examined each of the topics individually to understand the underlying sentiment tones of users, which turned out to be overwhelmingly negative for all topics, as depicted in Fig. 7 (data normalized to be presented as percentage) and negative in general across the three years. Some topics have a relatively higher percentage of tweets with a 'strongly positive' score, e.g., tweets belonging to the topics 'VPNuse' and 'HomePrivSec'. By manually inspecting some tweets under these topics, we noticed that many tweets are about suggesting devices or apps that users felt happy about, and some are about users expressing appreciation on being able to understand the workings of a technology or to remember their forgotten passwords. Two example topics tweets with a negative sentiment talked about are Wi-Fi access points being hacked by neighbors and vulnerabilities of PMs. Two topics, 'InternetSec' and 'LaptopSec', have the higher percentage (over 70% across all three years) of tweets with a 'strongly negative' score. Our results showed a sharp contrast to the results reported in (Sriram et al., 2021), which studied Twitter data to determine users' sentiment towards IoT in general and found the users sentiments to be more positive towards cyber security and privacy.

## 5. Further discussions, limitations and future work

To summarize the results reported in the previous section, the good performance of the two classifiers enabled us to extract a large number of cyber security and privacy-related tweets posted by non-expert users in three consecutive years (January and February of 2019-21). Topical modeling analysis of the extracted tweets led to insights about a wide range of topics non-expert users were discussing before the global COVID-19 lockdowns and during the pandemic. The biggest topic is about the use of VPNs, which was unexpected because the topic has not been extensively studied in the research literature (more details are given below). Comparing topics in 2019-20 and 2021, we noticed a substantial increase in non-expert users' discussions on many topics including VPN, laptops, mobile phones, and device accounts. Sentiment analysis of the extracted tweets led to the discovery of an overall and consistent negative sentiment for all topics across all three years. However, we found that non-expert users' opinions seemed to have become more polarized during the COVID-19 pandemic (2021), i.e., the neutral sentiment became much thinner compared with before the global lockdowns (2019-20). Another noticeable pattern we observed is that most discussions of non-expert users were about traditional computing devices such as laptops, tablets, and smartphones rather than smart home devices, **suggesting that over-focusing on cyber security and privacy of smart home devices may miss the likely fact that traditional computing devices remain the center of home networking at most households.**

### 5.1. Need of research on non-technical aspects of the use of VPNs

The fact that VPN-related discussions form the biggest topic in our data can be related to the reported increased usage of VPNs, especially during the COVID-19 pandemic (Johnson, 2021b). Among all such discussions about VPNs, typical sub-topics include privacy, malware, the legality of use, bypassing geolocation control (based on IP addresses), hardware- and software-based VPNs, and work-based tweets. The extensive discussions on VPNs are aligned with the results reported in (Busse et al., 2019), which reported that VPNs are often mentioned in cyber security and privacy advice and among general practices of non-expert users. Despite the importance of VPNs, current research on VPNs is more about technical aspects, and human factors are much less studied. Studies that did venture into human factors of VPNs focused on limited areas such as the information disclosed while using VPNs (Molina et al., 2019), user awareness of VPN (Jayatilleke and Pathirana, 2018; Karaymeh et al., 2019), and user attitudes towards adopting VPNs (Namara et al., 2020; Sombatruang et al., 2020) (note that (Namara et al., 2020) focused mainly on expert users). We believe more research is needed to study a **wide range of aspects regarding the use of VPNs by non-expert users, especially in complicated contexts such as working from home, bringing own devices to work, and other hybrids (e.g., work-life, multi-user and multi-device) environments**.

*5.2. Meeting the cyber security help needs of non-expert users*

Another noticeable phenomenon we observed is that many discussions of non-expert users, either about traditional computing devices, smart devices, or other digital platforms and tools, were often about seeking help and advice and expressing frustration about cyber security and privacy. Such discussions formed one of the topics in 2021 (during the COVID-19 pandemic), indicating that working and studying from home led to an increased level of cyber security and privacy concerns and problems. Many of these tweets, have negative sentiment, suggesting that non-expert users largely struggled with getting their problems solved without help. According to our analysis, non-expert users often referred to Google for support on cyber security and privacy problems, however, as reported in a recent study (Turner et al., 2021), googling often did not work appropriately for getting such advice. We call for more research on **understanding difficulties non-expert users are facing with getting the help they need and on better methods for providing such help**.

*5.3. Cyber security and privacy related issues from the connected home networking ecosystem with multiple devices and multiple users*

A third key finding of our study is about non-expert users' discussions on their usage of multiple devices and their interactions with multiple other users in the home context. Some discussions on multiple devices are less surprising, e.g., interactions between mobile phones and smart devices, but some others were less expected, e.g., security issues of Bluetooth pairing between different computing devices. Regarding multi-user aspects, we noticed discussions that pointed to scenarios where cyber security and privacy issues were caused by or related to the presence of multiple users at home. Example scenarios include non-expert users sharing their Wi-Fi access points and password details with secondary and temporary users such as house guests and friends, and some users discussing how to illegally break into neighbors' Wi-Fi access points. Past studies on such multi-user aspects we are aware of (Ahmad et al., 2020; Bernd et al., 2020; Huang et al., 2020b; Marky et al., 2020) were all smaller-scale empirical studies. One possible future research direction on **multi-user and multi-device scenarios is to investigate how personal and sensitive data flow between different computing devices and possible users.** The subject of personal and/or sensitive data is an important topic not only because it is an element in understanding and maintaining the privacy of data, but also because the sensitivity of the personal data alters depending on the context it is used, the source it is derived from and the stakeholder which uses it and evolve with time (Saglam et al., 2022). A detailed study of these areas can help understand user behaviors and identify new interventions.

*5.4. Limitations*

Our study demonstrated that using real-world data from OSN platforms can offer a much richer dataset and cover a wider range of topics discussed by people in the wild than empirical studies based on a smaller number of participants in settings such as surveys, interviews, and focus groups. However, using online social media analysis along also has its own limitations, e.g., it is more passive and depends on what people discussed online. There could be more issues at play in the real world that would not be captured through online postings. It is also difficult to study more specified topics or specific interventions when studying the sheer volume of data. The diversity of the data also means there can be too many factors affecting the results, so it can be difficult to conduct causal analysis. As a result, we believe that mixed methods, which combine both larger-scale OSN analysis and smaller-scale empirical studies, will be more appropriate for studying user perspectives on cyber security and privacy. For instance, as suggested in (Joseph et al., 2021), the larger set of macro results obtained via an OSN analysis study could be followed by complementary empirical studies that focus on more specific aspects revealed by the OSN analysis.

Secondly, we have used the months of January and February of each year to conduct this study, which might not be representative enough of the whole year. In future studies, we would endeavor to include a more representative set of data to give us more informative results and coverage.

Last but not the least, we would like to point out that our study is heavily based on the two machine learning classifiers and the automated analysis tools for topical modeling and sentiment analysis. Such tools are not perfect and can produce both false positives and false negatives, and the high volume of data made a manual inspection of results infeasible, so we relied on a limited level of qualitative analysis. The manual labeling process we did could also suffer from subjectivity, since we had to rely on some subjective judgments when applying pre-defined criteria to label the data. The keywords we used to extract the original 13.7 million tweets may not be representative enough, so we may have missed some relevant tweets. All such problems are common for all online social media analysis studies, which further justifies the need to have some follow-up smaller-scale empirical studies to validate and extend the findings reported in our study.

## 6. Conclusion

Using a large-scale Twitter dataset in January and February in three consecutive years (2019-21) and two BERT-based classifiers to automatically detect cyber security and privacy-related tweets posted by non-expert users, this paper reports interesting results about non-expert users' discussions on cyber security and privacy, via topic modeling and sentiment analysis. Past studies have used classifiers to filter cyber security related tweets, but this study focused on cyber security and privacy related tweets posted by non-expert users and investigated the identified tweets both qualitatively and quantitatively. Our analysis comprehensively covers three different perspectives, i.e., overall analysis across the three years covered, yearly analysis, and a comparative analysis of tweets posted before and during the COVID-19 pandemic.

We observed that non-expert users discussed a wide range of topics, covering many different types of computing devices, with more discussions on more traditional computing devices than smart ones. The results also showed noticeable changes in non-expert users' discussions on cyber security and privacy before the global COVID-19 lockdowns (2019-20) and during the pandemic (2021), e.g., an increased level of more polarized sentiment in 2021, a focus on help-related tweets relating to cyber security and privacy, and the increased use of specific cyber security tools such as VPNs. The study endeavours to inform researchers about various topics for future research, e.g., more research on the use of VPNs by non-expert users, helping non-expert users on cyber security and privacy, and cyber security and privacy issues arising from the use of multiple devices in the home without ignoring traditional devices, and cyber security and privacy risks related to different roles and interactions of multiple users of a home network. These topics should be studied via more data-driven research and also more diverse empirical studies.

**Declaration of Competing Interest**

The authors declare that they have no known competing financial interests or personal relationships that could have appeared to influence the work reported in this paper.

**CRediT authorship contribution statement**

**Nandita Pattnaik:** Conceptualization, Methodology, Software, Data curation, Formal analysis, Writing – original draft. **Shujun Li:** Conceptualization, Methodology, Formal analysis, Writing – review & editing. **Jason R.C. Nurse:** Conceptualization, Methodology, Writing – review & editing.

**Data Availability**

Data will be made available on request.

**Supplementary material**

Supplementary material associated with this article can be found, in the online version, at doi:10.1016/j.cose.2022.103008.

**Nandita Pattnaik** is a Ph.D. student in Computer Science at the University of Kent and a member of the Institute of Cyber Security for Society (iCSS). Her research focuses on cyber security and privacy awareness, behaviour, practices and concerns in multi-user homes, where the inhabitants use multiple smart and traditional computing devices. Nandita has been teaching Computer Science to higher secondary and undergraduate students for most parts of her career in various parts of the world, including India, Oman and the UK. She has worked as a successful corporate trainer, planning, developing, and delivering customised training on specialised computing software to small businesses and big corporations in India and UK. Nandita holds an honours degree in Analytical Economics and a BSc. In Computer Science.

**Shujun Li** is Professor of Cyber Security at the School of Computing and Director of the Institute of Cyber Security for Society (iCSS), University of Kent, UK. He has published over 100 scientific papers, including five Best Papers including the 2022 IEEE Transactions on Circuits and Systems Guillemin-Cauer Best Paper Award. His research interests are about inter-disciplinary topics related to cyber security and privacy, human factors, digital forensics and cybercrime, multimedia computing, and data science. Professor Shujun Li is on the editorial boards of multiple international journals and participated in the organisation of more than 100 international conferences and workshops. He is a Fellow of the BCS, a Senior Member of the IEEE, a Professional Member of the ACM, and a Vice President of the Association of British Chinese Professors (ABCP).

**Dr Jason R. C. Nurse** is an Associate Professor in Cyber Security in the School of Computing at the University of Kent, UK and the Institute of Cyber Security for Society (iCSS), UK. He also holds the roles of Visiting Academic at the University of Oxford, Visiting Fellow in Defence & Security at Cranfield University, UK and Associate Fellow at the Royal United Services Institute for Defence and Security Studies (RUSI). His research interests include security and privacy risk management, corporate communications and cyber security, secure and trustworthy Internet of Things, insider threat and cybercrime. Jason was selected as a Rising Star for his research into cybersecurity, as a part of the UK's Engineering and Physical Sciences Research Council's Recognising Inspirational Scientists and Engineers (RISE) awards campaign. Dr Nurse holds a Ph.D. in cyber security, an M.Sc. in Internet Computing and a B.Sc. in Computer Science and Accounting. He has published over 100 peer-reviewed articles in internationally recognised security journals and conferences.